# Effect of Lockdown on the spread of COVID-19 in Pakistan

Fizza Farooq[1], Javeria khan[1], Muhammad Usman Ghani Khan[1,2]

[1]Al-Khawarizmi Institute of Computer Science (KICS), UET Lahore

[2]Department of Computer Science, UET Lahore

**Abstract**

A novel coronavirus originated from Wuhan, China in late December 2019 has now affected almost all countries worldwide. Pakistan reported its first case in late February. The country went to lockdown after three weeks since the first case, when the total number of cases were over 880. Pakistan imposed a lockdown for more than a month which slowed the spread of COVID 19 effectively, however in late April relaxation in lockdown was allowed by the government in stages to lift the strain on the economy. In this study, the data has been analyzed from daily situation reports by the National Institute of Health Pakistan and the effects of initial strict lockdown and later smart lockdown have been studied. Our analysis showed a 13.14 Percentage increase in cases before lockdown which drops down to 6.55 percent during the lockdown. It proved the effectiveness of lockdown. However, the Percentage Increase in case grows up to 7.24 during a smart lockdown. If it continues to rise in this manner, Pakistan may need to enter again into a strict second lockdown.

## Introduction

COVID-19, previously named as 2019 novel coronavirus (2019-nCoV) is a severe acute respiratory syndrome caused by a virus named as SARS-CoV-2. It is the third coronavirus disease after SARS-CoV in 2002 to 2003 and Middle East respiratory syndrome coronavirus (MERS-CoV) in 2012. Both SARS and MERS have a high fatality rate than SARS-CoV-2 but less spread rate. At the end of the SARS epidemic in 2003, there were 8098 people infected worldwide and 774 died with a 9.6% fatality rate[1]. By November 2019, 2494 confirmed cases of MERS have been reported with 858 deaths and a fatality rate of 34.4%[2]. On the other hand COVID-19, according to WHO situation report of May 4, 2020, has 3,435,894 confirmed cases with 239,604 deaths (fatality rate of 6.9%) worldwide[3].

COVID-19 originated in Wuhan, Hubei province of China in late December 2019, when some patients with an initial diagnosis of pneumonia of unknown etiology were admitted to hospitals. These patients were epidemiologically linked to the seafood and wet animal wholesale market in Wuhan[4][5]. As of January 22, 2020, a total of 571 cases of COVID-19 were reported in 25 provinces in China[6]. On January 30, 2020 WHO declared it as Public Health Emergency of International Concern (PHEIC) when the virus has spread to 18 countries outside China with a total of 7818 cases worldwide[7]. On March 11, 2020 when the disease has spread to 114 countries with more than 118,000 confirmed cases and over 4200 deaths, WHO declared it a global pandemic[8].

Pakistan reported the first two cases of COVID-19 on 26 February 2020. Both patients had a travel history of Iran[9]. Within two weeks the number of cases reached up to 20 and all have travel history of Iran, China, Syria, and London[10]. By 23 March the total number of cases was 892 with 6 causalities. With no vaccine currently available, the only solution is to prevent and contain the disease by lockdown and social distancing. The countrywide lockdown was enforced on 24 March which continued for more than a month. A major cause of the initial spread of the virus in Pakistan is pilgrims returning from Iran at Taftan border. Over 7000 pilgrims returned from Iran out of which 1433 have been tested positive according to the situation report by NIH on 4 May[11]. Another cause of rapid spread is the religious mass gathering at Raiwind, Lahore in early March. About 80000 to 1250000 people participated out to which 3000 were foreigners from 40 different countries[12]. By 4 May, over 3033 cases had a history of attending this gathering. On 11 May 2020, a total number of 30941 cases have been reported with 8212 recoveries and 667 deaths[13].

As the vaccine of COVID-19 is not currently available, the only solution to handle this outbreak is to prevent its spread and take effective measures to contain it. More than 100 countries had to go into either full or partial lockdown to contain this outbreak[14]. China went into partial lockdown after three weeks of the first reported case and lasted till mid of March. Iran was a hard-hit country in Middle-East Asia by this outbreak. The first cases were reported in late February and more than 107 thousand people have been affected by it in Iran. Iran went into complete lockdown in early March. Pakistan, India, Saudi Arabia, Turkey, and Kuwait also practiced massive lockdowns. Europe became the epicenter of the outbreak after China. Italy and Spain were the worst-hit countries with over 219,070 and 224,350 confirmed cases respectively. Both countries imposed initially partial lockdown but later more restrictive and extensive lockdowns which helped in slowing down the spread[15].

The objective of this study is to highlight the effects of lockdown on the spread of COVID-19 in Pakistan. The government did not impose full lockdown abruptly in the country instead it was imposed gradually. The statistical analysis of COVID-19 data has been performed depending upon the actions taken by the government and the timeline of lockdown has been into three categories. Finally, the impact of each category on the growth rate of the disease in Pakistan has been studied.

**Methodology**

We used the dataset of Pakistan available at Kaggle[16]. The source of this data is COVID-19 daily situation report generated by the National Institute of Health (NIH) Pakistan from 10 March 2020 to 9 May 2020. The lockdown in Pakistan was not enforced abruptly but gradually in different stages. Based on these stages we divided the whole timeline into three phases. Phase A is before actual lockdown, phase B is a full lockdown, and phase C is smart lockdown or relaxation in lockdown. Table 1 shows the steps taken by the government of Pakistan gradually towards lockdown and smart lockdown. We presented data in a way to get insight into the trend of the spread of COVID-19 during different phases of lockdown in Pakistan. We calculated the average rise of confirmed cases in each phase. We also calculated the average value of Percentage Increase in each phase from daily incident data.

Table 1: Actions that are taken in different stages of lockdown by the government of Pakistan

| Phase A<br>26 Feb – 23 March | Phase B<br>23 March – 25 April | Phase C<br>25 April onwards |
|---|---|---|
| • Closed Pak-Iran border until 7 March and suspend all flights from Iran<br>• Ensure screening and quarantine facility for pilgrims returning from Iran before re-opening Taftan border<br>• Thermal screening at 4 major airports<br>• Public awareness regarding COVID-19<br>• Section 144 imposed to ban all public gatherings, restaurants and marriage ceremonies<br>• All educational institutions are closed<br>• All international flights to be suspended till 4 April | • All industries, businesses, offices were closed<br>• All malls and markets were also closed<br>• Grocery and pharmacy shops open for limited hours during the day<br>• All public transports were stopped<br>• Work from home<br>• Online classes and lectures started<br>• Only essential supply chain services related to food and medicine were allowed | • Low-risk industries and businesses were allowed to re-open following SOP, social distancing among employees and ensuring workplace cleanliness<br>• Markets to open for a few hours in a day<br>• Restaurants and malls are not allowed to open<br>• Food delivery services re-opened<br>• Public transport cannot be started |

**Results**

A constantly increasing trend in the daily confirmed cases can be seen in figure 1. The relatively slow spread in Phase A can be justified by a smaller number of testing capacity and facility as the disease has just started in Pakistan. The actual number of infected patients in Phase A must be higher as reported. Pakistan imported PCR based testing kits mostly from China and extensive testing started from Phase B.

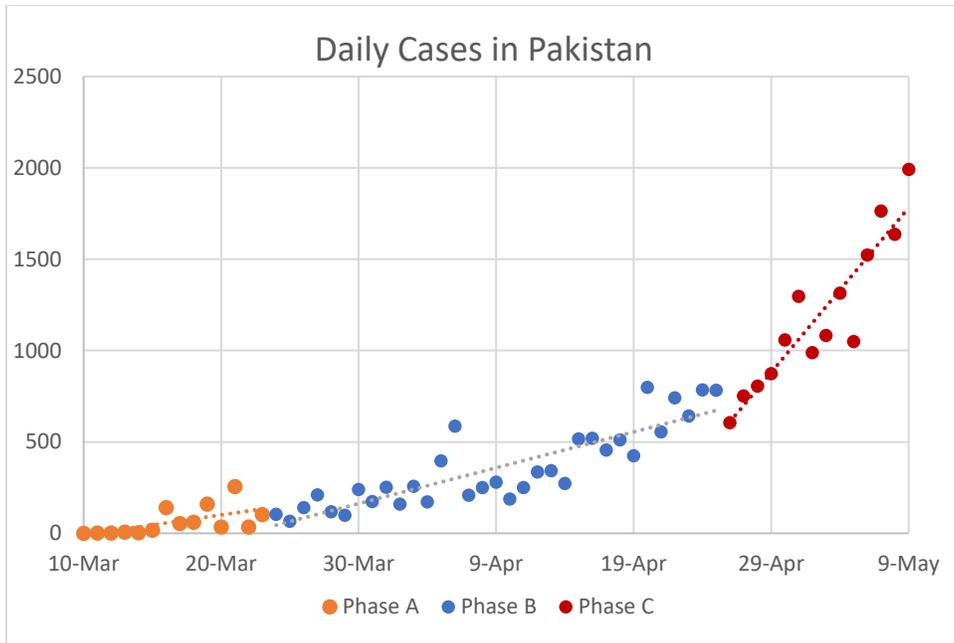

*Figure 1 Daily cases in Pakistan and their trend*

A linear trend is calculated and displayed for each phase. There is a sudden increase in trend in Phase C as an effect of relaxation in lockdown. Figure 2 shows that on an average approximately 67 new cases were reported daily in phase A, 358 in phase B while 1195 in phase C. The highest number of new cases were reported on 9 May 2020 which is 1991 as shown in figure 3.

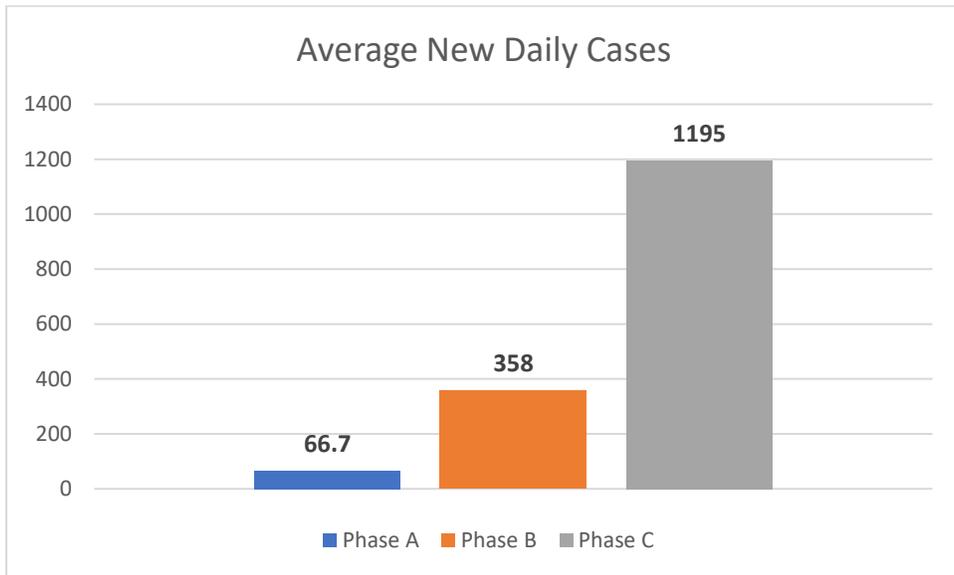

*Figure 2 Average number of new cases in different phases of lockdown*

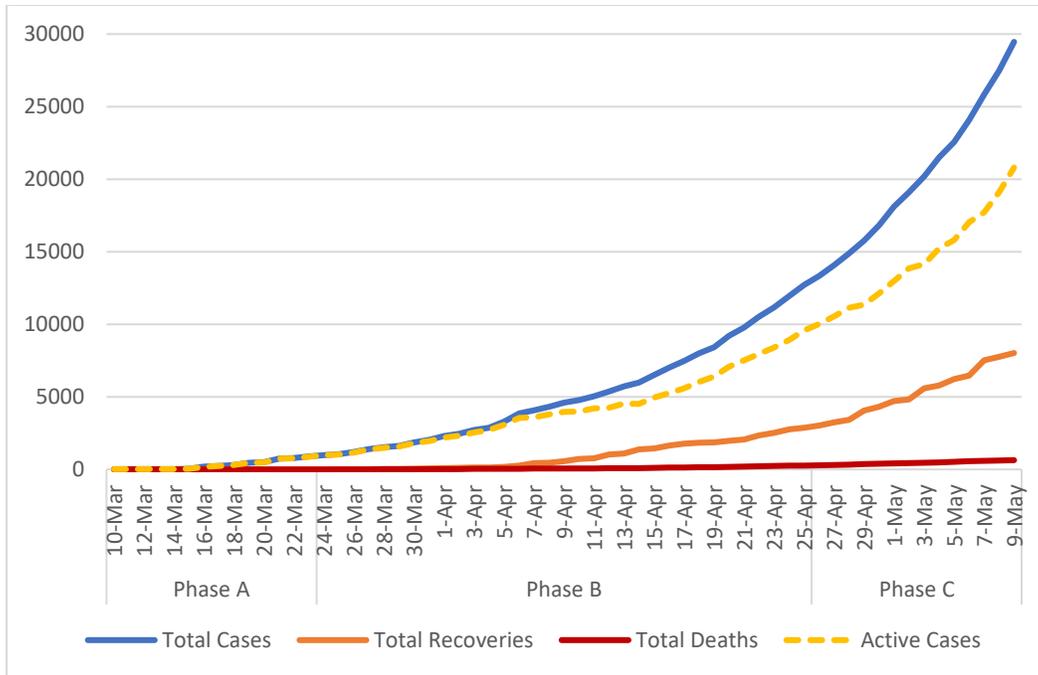

Figure 3 Overview of COVID-19 in Pakistan

Figure 4 shows the Percentage Increase in the number of cases in each phase. Before lockdown it was 13.14%, during lockdown it decreased significantly to 6.55% (-6.59% drop) and in smart lockdown it goes up to 7.24% (+0.69% rise).

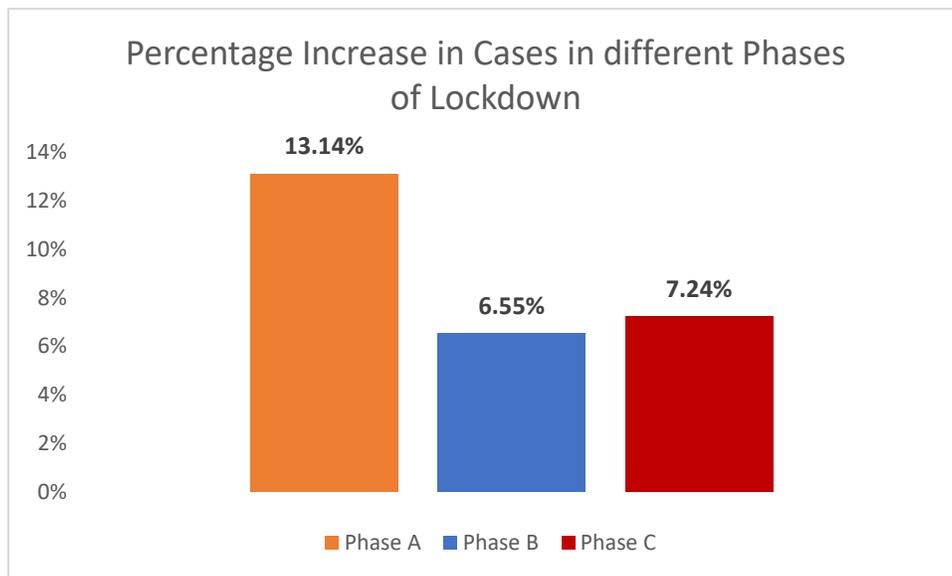

Figure 4 Percentage Increase in cases in different phases

**Discussion**

The strict lockdown which we named phase B, showed a relatively slow, steady, and controlled spread of COVID-19 quite as expected. It was fulfilling its purpose effectively towards the control of pandemic in the country but on the other hand it was causing difficulties for daily wagers, and small businesses. The complete strict lockdown continued for more than a month. It affected the country's economy as all industries and import exports were ceased. For a developing country like Pakistan, it was difficult to keep strict lockdown for a long time. Hence, the government announced a smart lockdown to be enforced in stages. It means to re-open low-risk industries like manufacturing businesses, construction, industries related to food and agriculture, and factories with daily wagers and labor. All businesses that were re-opened were directed to follow SOP regarding workplace cleanliness, use of hand sanitizers and masks, and maintaining social distance. After this relaxation in the lockdown, the number of new daily cases increased abruptly. An increase in free-roaming and unnecessary movement of people was also noticed which should be handled strictly by special forces in every district. If this situation is not controlled then in a very short time it could be unmanageable and our health facilities could not capacitate them. There may be a need for a second lockdown for that which could have even worse effects on the economy.

**Conclusion**

The lockdown from 23 March 2020 to 25 April 2020 proved quite effective in slowing down the growth of the outbreak in Pakistan. During this period, the number of daily reported cases drop by 6.55%. Due to the economic condition of the country, the government allowed some low-risk businesses to re-open, and hence mobility of people increased causing a 0.69% rise in daily reported cases. There is an immense need of following strictly to the guidelines provided by NIH for all re-opening businesses. The general public should also abide by the rules of home quarantine, hygiene, and all preventive measures. If the situation deteriorates then Pakistan might need to enter again into full second lockdown.